\newcommand\aj{{AJ\,}}%
\newcommand\apj{{ApJ\,}}%
\newcommand\apjs{{ApJS\,}}%
\newcommand\aap{{A\&A\,}}%
\newcommand\pasp{{PASP\,}}%
\documentclass[graybox, natbib, footinfo]{svmult}

\usepackage{mathptmx}       % selects Times Roman as basic font
\usepackage{helvet}         % selects Helvetica as sans-serif font
\usepackage{courier}        % selects Courier as typewriter font
\usepackage{type1cm}        % activate if the above 3 fonts are
\usepackage{makeidx}         % allows index generation
\usepackage{graphicx}        % standard LaTeX graphics tool
\usepackage{multicol}        % used for the two-column index
\usepackage[bottom]{footmisc}% places footnotes at page bottom

\makeindex             % used for the subject index
\begin{document}

\title*{The APOKASC Catalog}
% Use \titlerunning{Short Title} for an abbreviated version of
% your contribution title if the original one is too long
\author{Jennifer A. Johnson}
% Use \authorrunning{Short Title} for an abbreviated version of
% your contribution title if the original one is too long
\institute{Jennifer A. Johnson \at Department of Astronomy, Ohio State University, 140 W. 18th Ave, Columbus, OH, 43210 \email{jaj@astronomy.ohio-state.edu}}
%\and Name of Second Author \at Name, Address of Institute \email{name@email.add%ress}}
%
% Use the package "url.sty" to avoid
% problems with special characters
% used in your e-mail or web address
%
\maketitle

\abstract{
I report on the APOKASC catalog, a joint effort between the
  {\it Kepler} Asteroseismic Science Consortium and the SDSS-III APOGEE spectroscopic
  survey. It will contain both seismic and spectroscopic values for
stars observed by both surveys. I discuss the derivation of 
spectroscopic parameters and their uncertainties from the H-band spectra
delivered by the APOGEE spectrograph, illustrating the sensitivity
of stellar spectra to some parameters, such as T$_{\rm eff}$, and 
lack of sensitivity to others, such as logg.}

\section{Contributors to the APOKASC catalog}
\label{sec:1}
This catalog is the result of intensive effort by KASC and APOGEE
scientists, particularly those who have been working on the analysis
of {\it Kepler} lightcurve data, grid-based modelling, and the APOGEE
spectroscopic parameters pipeline (ASPCAP), but also the
broader teams from the instrument builders to lead scientists. I
acknowledge here in reverse alphabetical order the former group: Gail
Zasowski, Nicholas Troup, Dennis Stello, Verne Smith, Victor Silva
Aguirre, Matthew Shetrone, Aldo Serenelli, Ricardo Schiavon, Marc
Pinsonneault, Benoit Mosser, Andrea Miglio, Travis Metcalfe, Szabolcs M{\'e}sz{\'a}ros, 
Savita Mathur, Steven R. Majewski, Daniel Huber, Jon Holtzman, Saskia
Hekker, Leo Girardi, Rafael Garcia, Ana Garc{\'i}a P{\'e}rez, 
Courtney R. Epstein, Yvonne
Elsworth, Katia Cunha, William Chaplin, Dmitry Bizyaev, Sarbani Basu,
Carlos Allende Prieto.

\section{Overview of APOKASC Catalog}

The APOKASC catalog is a joint effort between the
  {\it Kepler} Asteroseismic Science Consortium and the SDSS-III APOGEE spectroscopic
  survey.  For stars observed by both groups, the catalog will have
  the seismic parameters $\Delta\nu$ and $\nu_{\rm max}$, the
  spectroscopic parameters T$_{\rm eff}$ metallicity and their
  uncertainties. In addition, grid-based modeling values for logg,
  mass and radius, derived using the combination of seismic and
  spectroscopic parameters, will be reported. The first version of the
  catalog  will contain the results for
$\sim 1900$ red giant stars \citep{Pinsonneault}; subsequent versions of the catalog will include
thousands of additional red giants observed in Years 2-3 of APOGEE
observations as well as hundreds of dwarf stars.

\section{Advantages of Combining Spectroscopic and 
Seismic Data }
\label{sec:2}

Measurements of the fundamental properties of stars, such as masses,
radii, ages, rotation profiles, evolutionary state, and
composition, are crucial for understanding numerous issues in
stellar structure and evolution, star formation histories, and
stellar populations. Seismic measurements and spectroscopic measurement
are very complementary in which characteristics they can measure.
For example, logg is very well determined from the seismic
parameters, while detailed composition measurements, such
as [Mg/Fe] or [Na/Fe], will remain the purview of spectroscopy.
Also, the addition of T$_{\rm eff}$ and metallicity information can 
improve the accuracy of grid-based modelling results for properties
such as mass (e.g., \citealt{Gai}).

With this catalog, we plan to study a variety of areas, including
the projects below, which have already been started from the
preliminary version of the APOKASC catalog.

\begin{itemize}
\item{Dependence of seismic properties on metallicity}
\item{Surface and internal stellar rotation rates}
\item{New identifications of stellar populations, e.g., merger remnants}
\item{Improved distance measurements}
\item{Tests of Galactic stellar population models from masses and radii of
thousands of red giants}
\end{itemize}

\section{Spectroscopic Data }
\label{sec:3}
% Always give a unique label
% and use \ref{<label>} for cross-references
% and \cite{<label>} for bibliographic references
% use \sectionmark{}
% to alter or adjust the section heading in the running head

I focus this proceeding on describing the APOGEE spectroscopic observations
and the sensitivity of near-IR spectra to stellar parameters, as this
audience is most familar with the derivation and use of seismic parameters.

\subsection{Overview of APOGEE survey}
\label{subsec:2}
The Sloan Digital Sky Survey (SDSS) III (\citealt{Eisenstein}) uses a
2.5-meter telescope with a wide field of view (7 square degrees) 
at Apache Point Observatory (APO) to survey the sky. The APO
Galactic Evolution Experiment (APOGEE) uses a fiber-fed
multiobject near-infrared spectrograph to observe 230 science targets
at a time \citep{Majewski}. An additional 35 fibers are reserved for observing
hot stars to facilitate removal of telluric lines and
35 fibers are placed on regions of black sky. The spectra
are high resolution (R$\sim$22,500), high signal-to-noise ($\sim$ 100 per
dithered resolution element) and cover the wavelength range
of 1.51 to 1.68$\mu$m. APOGEE began routine observations
in Fall 2012. The first data release of APOGEE data, including
the spectra for the stars that are part of the first APOKASC
catalog, was the SDSS Data Release 10 \citep{Ahn}.

\subsection{Derivation of Stellar Parameters from Spectra }

The strengths of absorption features in the spectra of cool giants
are sensitive to several properties of the atmosphere, including
T$_{\rm eff}$ and logg, which affect the excitation and ionization
balance of the electrons and the molecular equilibrium, 
and compositon, which affects both the optical depth at the 
wavelengths of transitions and the continuous opacity through electron
donation to form H$^-$.

To derive the stellar atomspheric parameters from spectra, first, 
the raw data are processed through the data reduction
pipeline to produce combined, wavelength-calibrated, and 
radial-velocity-corrected spectra \citep{Nidever}.
Next, stellar parameters are determined using ASPCAP \citep{Garcia},
which finds the $\chi^2$ minimum difference between the observed
spectra and a 6-D grid of synthetic spectra. The parameters
that are varied to create the synthetic spectra are 
T$_{\rm eff}$, logg, [M/H], [C/M], [N/M] and [$\alpha$/M]. To
calculate spectra with different [M/H], the abundances of all
elements except C, N, and the $\alpha$ elements are changed up
or down in the same proportion from their solar ratios. In as
much as lines of Fe are the most common out of all the elements
that comprise ``M'', the best-fit [M/H] is an excellent proxy for
[Fe/H], a fact confirmed by comparison with cluster stars \citep{Meszaros}.
Fitting to a 6-D grid is necessary because the depth of an 
absorption line is never sensitive to just one parameter.
A critical assesment of the systematic and random uncertainties
in the ASPCAP stellar parameters was performed by comparison
with parameters for stars in well-studied globular and
open clusters and for stars in the ``gold standard'' KASC
seismic sample in the {\it Kepler} field \citep{Meszaros}. 

\subsubsection{Effective Temperature}

\begin{figure}[h]
%\sidecaption
% Use the relevant command for your figure-insertion program
% to insert the figure file.
% For example, with the graphicx style use
\includegraphics[angle=270,scale=.48]{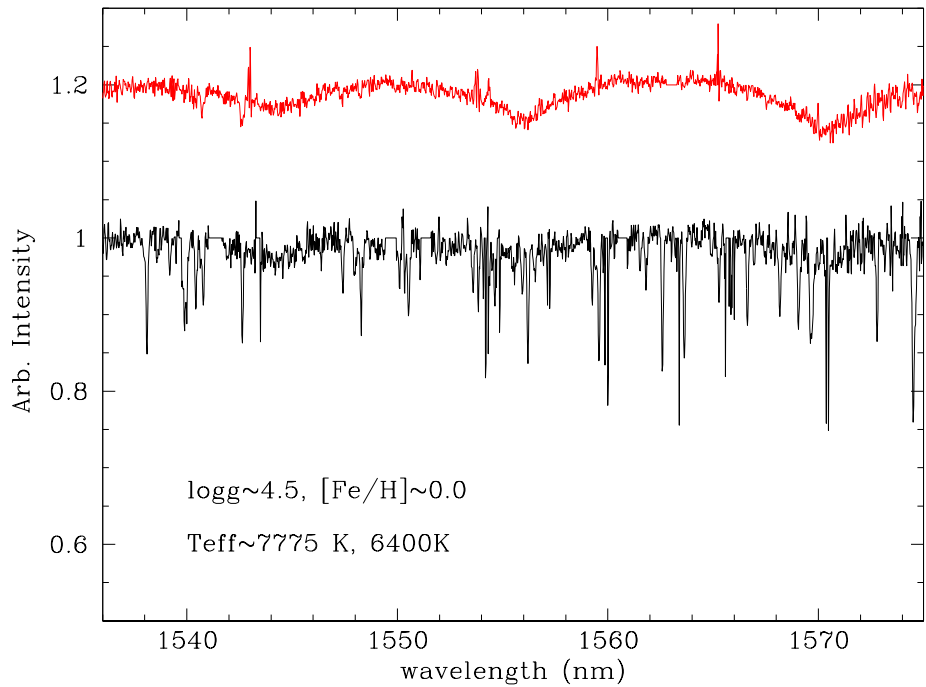}
%
% If no graphics program available, insert a blank space i.e. use
%\picplace{5cm}{2cm} % Give the correct figure height and width in cm
%
\caption{This figure shows parts of the spectra of two stars
in the APOKASC catalog that differ by $\sim$ 1400 K or $\sim$20\% in
T{\rm eff}, but have similar logg and [M/H]. 
Despite the relative small change in the temperature, there are
large differences in the appearance between the hotter star (top)
and the cooler star (bottom).}
\label{fig:1}       % Give a unique label
\end{figure}

The stellar parameter that has the most effect on the appearance
of a stellar spectrum is the T$_{\rm eff}$. 
Fig~\ref{fig:1} shows two stars in the APOKASC catalog whose
temperatures differ by a mere 20\%, yet the strength of the
Brackett lines varies markedly. There are numerous lines that
appear in the cooler star that are weak or non-existent in the
hotter star as either the excitation or the ionization state is
not appropriate for absorption of near-IR photons. The strong
dependence on temperature in the Boltzmann and Saha equations
and the presence of numerous lines of minority species, such
as neutral atoms and molecules, means that small temperature
uncertainties translate into larger uncertainties in [M/H].

To determine the accuracy of the ASPCAP-derived T$_{\rm eff}$, 
we compared the T$_{\rm eff}$ from the spectroscopic pipeline to 
photometric temperatures derived from the IRFM-based color-T$_{\rm eff}$
for 2MASS filters \citet{GHB} for cluster stars with well-determined
reddenings. The ASPCAP temperatures were $\sim$ 150K lower than the
photometric ones on average, with a clear trend with temperature.
A correction was applied to the ASPCAP temperatures to bring them
into better agreement with the latter values. After the correction,
the scatter in difference between the two measurements was $< 200$ K, 
dropping to $< 100$ K for stars with solar-like metallicity or
higher.

The advantage of determing a systematic correction to the ASPCAP
temperatures, rather than using photometric temperatures for all
APOGEE stars, is that a correctly calibrated spectroscopic temperature
is reliable even if the reddening is uncertain. Many APOGEE fields lie
in the midplane of the Galactic disk in regions of highly variable
extinction, and the reddening map in the {\it Kepler} field itself is
currently being revised \citep{Zasowski,Huber}.

\subsubsection{Metallicity}

\begin{figure}[t]
%\sidecaption
% Use the relevant command for your figure-insertion program
% to insert the figure file.
% For example, with the graphicx style use
\includegraphics[angle=270,scale=.48]{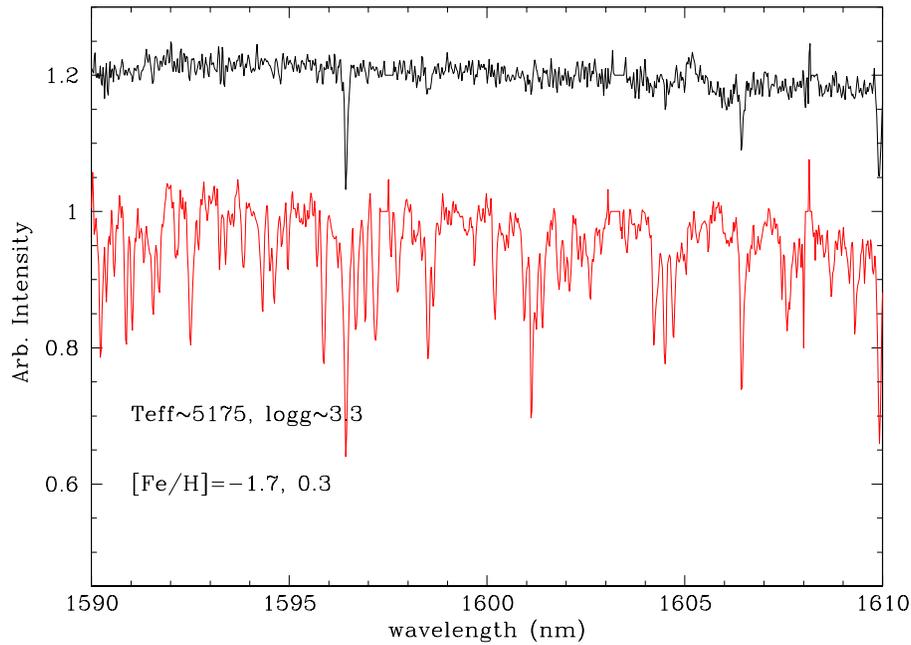}
%
% If no graphics program available, insert a blank space i.e. use
%\picplace{5cm}{2cm} % Give the correct figure height and width in cm
%
\caption{This figure shows parts of the spectra of two stars in the
the APOKASC catalog that differ by $\sim$ 2 dex in [M/H], but have
similar T$_{\rm eff}$ and logg. The number of heavy atoms in the
atmosphere has a clear effect on the spectrum, as the more metal-poor
star's spectrum (top) has considerably fewer lines than the more 
metal-rich star's spectrum (bottom). }
\label{fig:2}       % Give a unique label
\end{figure}

As Fig~\ref{fig:2} shows, the composition of a star has an obvious 
effect on the appearance of a spectrum, as more metal-rich stars
show much deeper absorption lines than a star with many fewer
heavy-element atoms in its atmosphere. The problem is, of course,
that temperature in particular has a much larger effect. The stars in
Fig~\ref{fig:1} differ in T$_{\rm eff}$ by $\sim$ 20\%, while the
two stars in Fig~\ref{fig:2} differ in metal content by a factor of
100! Therefore, the uncertainties in metallicity will be much
larger than the 4\% errors in T$_{\rm eff}$ discussed above.

The cluster stars provide the crucial
test of the uncertainties and systematic offsets for the ASPCAP
metallicity measurements, as all cluster stars should have
similar metallicity, at least at the level of precision
discussed here (e.g., \citealt{Carretta}), although subsequent
analysis will take into account the variation in globular
clusters of the light elements (e.g., \citealt{Kraft}). We compared
the [M/H] derived by ASPCAP to the [Fe/H] values reported in the
literature for clusters spanning a metallicity range of $-2.3 <$ [Fe/H] $< +0.3$
\citep{Meszaros}. In general, there was quite good agreement, confirming
that the [M/H] value is a good proxy for [Fe/H]. For metal-rich clusters,
the initial ASPCAP values are only about 0.1 too high, while they
are 0.35 dex for metal-poor clusters. A metallicity-dependent correction
is applied to the [M/H] values. The uncertainity in [M/H], 
determined by looking at the scatter in the corrected values compared
to literature values, ranges from 0.15 dex for the most metal-poor clusters
to $\sim$0.05 dex for clusters with solar metallicity and above. The
strong metallicity dependence in the uncertainty in metallicity
is not surprisingly, as metal-poor stars have noticeably fewer
metal lines to use for the measurement.

To assess the difference that the corrected and uncorrected
temperature scales and metallicity scales have on the results of the
grid-based modelling, two separate computations were done.
Uncorrected [M/H] were used in combination with uncorrected
T$_{\rm eff}$ values in one case, while corrected values are used in
the second case. Two sets of logg, masses and radii are reported.

\subsubsection{Gravity}

\begin{figure}[h]
%\sidecaption
% Use the relevant command for your figure-insertion program
% to insert the figure file.
% For example, with the graphicx style use
\includegraphics[angle=270,scale=.48]{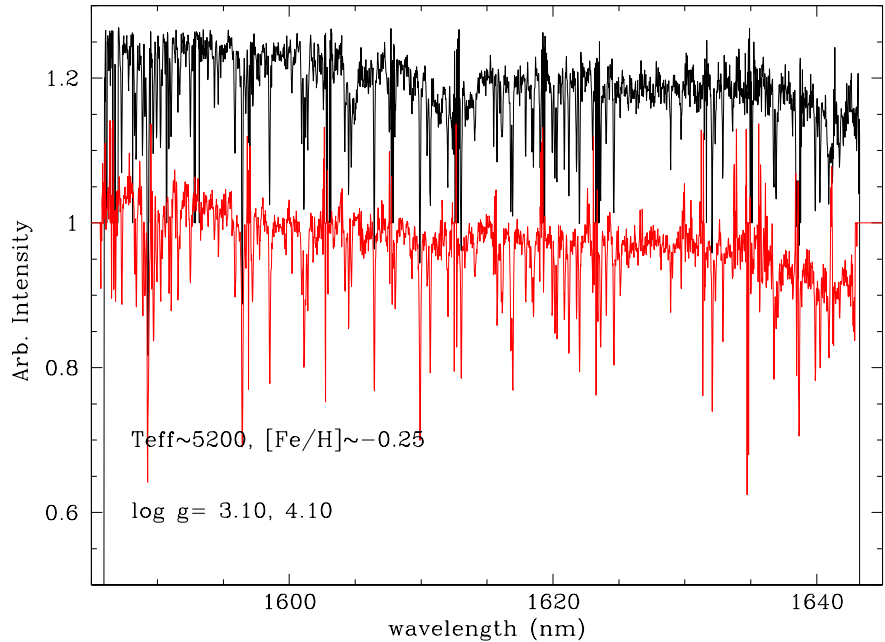}
%
% If no graphics program available, insert a blank space i.e. use
%\picplace{5cm}{2cm} % Give the correct figure height and width in cm
%
\caption{This figure shows parts of the spectra of two stars in the
APOKASC catalog that differ by $\sim$ 1 dex in logg, but have
similar T{\rm eff} and [M/H]. The differences
between the lower gravity star (top) and higher gravity star
(bottom) are not as obvious as for \ref{fig:1} and \ref{fig:2}, 
demonstrating why the uncertainties in gravity are large for
spectroscopically determined values.}
\label{fig:3}       % Give a unique label
\end{figure}

Logg values, unlike temperature and metallicity, can be measured by
both seismic and spectroscopic techniques. However, the spectroscopic
gravity measurements have larger uncertainties than the seismic
measurements because the spectra of stars in the H-band are not
particularly sensitive to logg changes (see Fig~\ref{fig:3}). We
compared the ASPCAP logg's with values both from seismic analysis of a
few hundred stars in the {\it Kepler} field as well as with the logg's
predicted for cluster stars from isochrones matched to the cluster
distance and reddening \citep{Meszaros}.  The values initially reported
by the ASPCAP pipeline are too high compared to the seismic and
isochrone stars by at least 0.2 dex, rising to 0.5 dex for metal-poor
stars. A correction is again applied, leading to much better
agreement. A scatter of $\sim 0.15$ dex is present in the differences
between corrected ASPCAP and calibration values, providing an estimate
of the remaining uncertainty in the spectroscopic values.  Part of
that scatter is because it was difficult to reconcile (at the $\sim$
0.1 dex level) the offsets between the seismic values in the {\it
  Kepler} field and the values determined from isochrones for stars
with similar T$_{\rm eff}$ and gravity in the clusters. This difference
is particularly striking for the metal-rich stars and remains under
investigation by the ASPCAP team.

% For figures use
%

\section{Conclusion}

In this proceeding, I have summarized the APOGEE data and the derivation
of stellar parameters and their uncertainties from high-resolution
H-band data. These parameters, along with the seismic parameters
$\Delta\nu$ and $\nu_{\rm max}$ are included in the APOKASC catalog 
as well as being passed to grid-based modellers for logg, mass, and
radii determinations that are also included in the catalog. The publication
of this catalog will, I hope, aid research in studies of stellar structure,
stellar atmospheres, and stellar populations.

\begin{acknowledgement}
A special thank you to Courtney Epstein, for her tireless work 
on targeting seismic sources in APOGEE and on gathering all the
data into the catalog. And thank you to everyone who measured
parameters for thousands of stars, no small task!
\end{acknowledgement}
%

%
%\section{References}
%References may be \textit{cited} in the text either by number (preferred) or by author/year.\footnote{Make sure that all references from the list are cited in the text. Those not cited should be moved to a separate \textit{Further Reading} section or chapter.} The reference list should ideally be \textit{sorted} in alphabetical order -- even if reference numbers are used for the their citation in the text. If there are several works by the same author, the following order should be used: 
%\begin{enumerate}
%\item all works by the author alone, ordered chronologically by year of publication
%\item all works by the author with a coauthor, ordered alphabetically by coauthor
%\item all works by the author with several coauthors, ordered chronologically by year of publication.
%\end{enumerate}
%You are encouraged to use BibTeX with the style \verb|spphys| provided with the template. Examples of citations are \cite{Kjeldsen95} and \cite{Reimers75}.

%\input{referenc}
% BibTeX users please use
%\bibliographystyle{spphys}
%\bibliography{author_sesto}

% Mac users: please ignore the error message: "! Package natbib Error: Bibliography not compatible with author-year citations."
\end{document}